\documentclass[aps,prb,preprint]{revtex4}%
\usepackage{amsfonts}
\usepackage{amsmath}
\usepackage{amssymb}
\usepackage{graphicx}%
\providecommand{\U}[1]{\protect\rule{.1in}{.1in}}
\begin{document}
\title[Maagnetic Moment Models]{Two Models Relevant to the Interaction of a Point Charge and a Magnetic Moment}
\author{Timothy H. Boyer}
\affiliation{Department of Physics, City College of the City University of New York, New
York, New York 10031}

\begin{abstract}
An understanding of the interaction of a point charge and a magnetic moment is
crucial for understanding the experiments involving electromagnetic momentum
carried by permeable materials as well as the experimentally-observed
Aharonov-Bohm and Aharonov-Casher phase shifts. \ Here we present two simple
models for a magnetic moment which have vastly different interactions with a
distant point charge. \ It is suggested that a satisfactory theoretical
understanding of the interaction is still lacking and that the
\textquotedblleft hidden momentum\textquotedblright\ interpretation has been
introduced into the textbook literature prematurely.

\end{abstract}
\maketitle

\section{Introduction}

The interaction of a point charge and a magnetic moment represents one of the
persistent problems of classical electromagnetism.\cite{Griffiths}$^{,}%
$\cite{Jackson} \ The interaction is intriguing partly because the combination
of the electric field of the charge and the magnetic field of the magnetic
moment introduces electromagnetic field linear momentum. \ One mystery of the
charge-magnetic-moment interaction is just how Nature incorporates this
electromagnetic field momentum into the relativistic conservation law for
linear momentum. \ The interaction of a point charge and a magnetic moment
forms the basis for controversies involving \textquotedblleft hidden
momentum,\textquotedblright\cite{Shockley}$^{,}$\cite{Vaidman}$^{,}%
$\cite{RLetter} the Aharonov-Bohm effect,\cite{AB} and the Aharonov-Casher
effect.\cite{AC} \ However, a certain version of the interaction involving
hidden momentum has made its way into the textbook literature\cite{Griffiths2}%
$^{,}$\cite{Jackson2} despite some objections.\cite{objection} \ Indeed some
research journals have rejected a competing description of the interaction,
even refusing to send out for review manuscripts which explore the competing
description.\cite{refuse} \ \ Most recently, the interaction has been used as
the basis for the astonishing claim that the Lorentz force law is incompatible
with relativity.\cite{Mansuripur} \ Given the controversies which still exist,
it seems wise to review the varying suggestions for the interaction of a point
charge and a magnetic moment. \ Here we will carry out calculations involving
two different models for a magnetic moment which interact with a point charge
in strikingly different ways. The contrasting behaviors suggest that there is
yet more to be understood regarding the interaction as it exists in Nature.
\ The contrast also suggests the possibility that the current
\textquotedblleft hidden momentum\textquotedblright\ interpretation in the
classical electromagnetism textbooks may be suspect and that the Aharonov-Bohm
effect and Aharonov-Casher effect may be wrongly presented in the quantum
mechanics texts.

\section{Two Models for a Magnetic Moment}

In the present article, two different models for a magnetic moment are
considered. \ The unperturbed magnetic-moment models both have a point charge
$e$ of mass $m$ moving in a circular orbit $\mathbf{r(}t)$ of radius $r_{0}$
and angular frequency $\omega_{0}$ about a charge $-e$ of large mass $M$
locate at the center of the orbit. \ Thus for the unperturbed magnetic moment
orbit, we have the displacement, velocity, and acceleration of the charge $e$
located at angle $\phi=\omega_{0}t+\phi_{0}$ given by%
\begin{equation}
\mathbf{r(}t)=r_{0}[\widehat{i}\cos(\omega_{0}t+\phi_{0})+\widehat{j}%
\sin(\omega_{0}t+\phi_{0})] \label{ee1}%
\end{equation}%
\begin{equation}
\mathbf{v}(t)=\omega_{0}r_{0}[-\widehat{i}\sin(\omega_{0}t+\phi_{0}%
)+\widehat{j}\cos(\omega_{0}t+\phi_{0})] \label{ee2}%
\end{equation}%
\begin{equation}
\mathbf{a}(t)=-\omega_{0}^{2}r_{0}[\widehat{i}\cos(\omega_{0}t+\phi
_{0})+\widehat{j}\sin(\omega_{0}t+\phi_{0})] \label{ee3}%
\end{equation}
The system has a magnetic moment
\begin{equation}
\overrightarrow{\mu}=\widehat{k}\frac{e\omega_{0}r_{0}^{2}}{2c} \label{ma1}%
\end{equation}
obtained from the ensemble-averaged (averaging over the initial phase
$\phi_{0})$ or time-averaged current density, and the steady-state current
formula $\overrightarrow{\mu}=%
%TCIMACRO{\dint }%
%BeginExpansion
{\displaystyle\int}
%EndExpansion
d^{3}r\,\mathbf{r}\times\mathbf{J}/(2c).$\cite{Jackson3} \ \ The crucial
difference between the models involves the binding which leads to the circular
orbit. \ The magnetic-moment model favored by the proponents of hidden
momentum involves a \textquotedblleft fixed-path\textquotedblright\ constraint
such that during any interaction the point charge $e$ may accelerate along the
circular orbital path but cannot depart from the circular orbital path of
radius $r_{0}$. \ The magnetic-moment model favored by the present author
involves a Coulomb-potential interaction between the charge $e$ and the
central charge $-e,$ corresponding to the behavior of a hydrogen-like atom.
\ Both these models have been presented previously in the literature, but
never before in direct comparison.\cite{comparison}

\section{Electromagnetic Momentum of Interaction}

We assume that the external charge $q$ (with which the magnetic moment is
interacting) is located on the $x$-axis at coordinate $x_{q},$ $\mathbf{r}%
_{q}=\widehat{i}x_{q},$ where $x_{q}>>r_{0},$ so that we may think of the
electric field $\mathbf{E}_{q}(\mathbf{r})=q(\mathbf{r}-\widehat{i}%
x_{q})/|\mathbf{r}-\widehat{i}x_{q}|^{3}$ due to the charge $q$ as
approximately constant over the magnetic moment at the value $\mathbf{E}%
_{q}=-\widehat{i}q/x_{q}^{2}$ with small corrections of order $r_{0}/x_{q}.$ \ 

The location of the magnetic moment in the electric field of the charge $q$
leads to linear momentum $%
%TCIMACRO{\dint }%
%BeginExpansion
{\displaystyle\int}
%EndExpansion
d^{3}r\mathbf{E}\times\mathbf{B}/(4\pi c)$ in the electromagnetic fields in
vacuum arising from the electric field $\mathbf{E}_{q}$ of the charge $q$ and
the magnetic field $\mathbf{B}_{e}$ of the magnetic moment. \ The
electromagnetic field momentum appears in order $v^{2}/c^{2},$ and therefore
we can work with the Darwin Lagrangian\cite{Darwin} and calculate
perturbations using nonrelativistic physics. \ Since the electromagnetic field
linear momentum $\mathbf{p}_{em}$ is already first order in the perturbing
charge $q$, the momentum\cite{PageandAdams} can be calculated through first
order in $q$ by using the unperturbed motion of the charge $e$ as%

\begin{align}
\mathbf{p}_{em}  &  =\frac{eq}{2c^{2}}\left(  \frac{\mathbf{v}(t)}%
{|\mathbf{r(t)-}\widehat{i}x_{q}|}+\frac{\mathbf{v}(t)\cdot(\mathbf{r(t)-}%
\widehat{i}x_{q})(\mathbf{r(t)-}\widehat{i}x_{q})}{|\mathbf{r(t)-}%
\widehat{i}x_{q}|^{3}}\right) \nonumber\\
&  =\frac{eq}{2c^{2}}\left[  \frac{\mathbf{v}(t)}{x_{q}}\left(  1+\frac
{\widehat{i}\cdot\mathbf{r}(t)}{x_{q}}\right)  +\frac{\mathbf{v}%
(t)\cdot\lbrack\mathbf{r(t)-}\widehat{i}x_{q}][\mathbf{r(t)-}\widehat{i}%
x_{q}]}{x_{q}{}^{3}}\left(  1+3\frac{\widehat{i}\cdot\mathbf{r}(t)}{x_{q}%
}\right)  \right]  \label{pp1}%
\end{align}
where we have expanded the denominators assuming that $r_{0}<<x_{q}$ to obtain
the expressions $|\mathbf{r(t)-}\widehat{i}x_{q}|^{-1}=x_{q}^{-1}%
(1+\widehat{i}\cdot\mathbf{r}(t)/x_{q}+...),$ $|\mathbf{r(t)-}\widehat{i}%
x_{q}|^{-3}=x_{q}^{-3}(1+3\widehat{i}\cdot\mathbf{r}(t)/x_{q}+...).$ \ Then
the average linear momentum $\left\langle \mathbf{p}_{em}\right\rangle $ in
the electromagnetic field is found from Eqs. (\ref{ee1}) and (\ref{ee2}) by
either ensemble averaging or averaging over time $t$ with $\left\langle
\cos^{2}\theta\right\rangle =\left\langle \sin^{2}\theta\right\rangle =1/2,$
$\left\langle \cos\theta\sin\theta\right\rangle =0,$%
\begin{equation}
\left\langle \mathbf{p}_{em}\right\rangle =\frac{eq}{2c^{2}}\left(
\widehat{j}\frac{\mathbf{\omega}_{0}r_{0}^{2}}{x_{q}^{2}}\right)  =-\frac
{1}{c}\overrightarrow{\mu}\times\mathbf{E} \label{pp2}%
\end{equation}
This is the electromagnetic field momentum which must be consistent with the
conservation of linear momentum.

\section{Nonrelativistic Perturbation Calculation -- Fixed-Path Model}

In the fixed-path model\cite{comparison} for a magnetic moment, the
current-carrying charge $e$ remains on the circular path of radius $r_{0}$ but
may change its velocity along this path. \ In the nonrelativistic perturbation
calculation, the angular acceleration $d^{2}\phi/dt^{2}$ of the charge $e$ in
its circular orbit is produced by the component of the electric field
$\mathbf{E}_{q}$ tangential to the orbit%
\begin{equation}
\frac{d^{2}\phi}{dt^{2}}=\frac{a_{\phi}}{r_{0}}=\frac{eE_{q}\sin(\phi)}%
{r_{0}m}=\frac{eq\sin(\phi)}{mr_{0}x_{q}^{2}} \label{ee4}%
\end{equation}
Since the angular acceleration is regarded as small, we may introduce the
unperturbed expression $\phi_{unperturbed}(t)=\omega_{0}t+\phi_{0}$ on the
right-hand side of Eq. (\ref{ee4}) to obtain%
\begin{equation}
\frac{d^{2}\phi}{dt^{2}}=\frac{eq\sin(\omega_{0}t+\phi_{0})}{mr_{0}x_{q}^{2}}
\label{ee5}%
\end{equation}
Assuming that the unperturbed motion holds at time $t=0,$ this equation can be
integrated with respect to time to obtain%
\begin{equation}
\phi(t)=\omega_{0}t+\phi_{0}-\frac{eq\sin(\omega_{0}t+\phi_{0})}{\omega
_{0}^{2}mr_{0}x_{q}^{2}} \label{ee6a}%
\end{equation}
Thus when interacting with the charge $q,$ the orbiting particle $e$ of the
magnetic moment has the position
\begin{align}
\mathbf{r}(t)  &  =r_{0}[\widehat{i}\cos(\phi)+\widehat{j}\sin(\phi
)]\nonumber\\
&  =r_{0}\left[  \widehat{i}\cos\left(  \omega_{0}t+\phi_{0}-\frac
{eq\sin(\omega_{0}t+\phi_{0})}{\omega_{0}^{2}mr_{0}x_{q}^{2}}\right)
+\widehat{j}\sin\left(  \omega_{0}t+\phi_{0}-\frac{eq\sin(\omega_{0}t+\phi
_{0})}{\omega_{0}^{2}mr_{0}x_{q}^{2}}\right)  \right] \nonumber\\
&  =r_{0}\widehat{i}\left[  \cos\left(  \omega_{0}t+\phi_{0}\right)
+\frac{eq\sin(\omega_{0}t+\phi_{0})}{\omega_{0}^{2}mr_{0}x_{q}^{2}}\sin
(\omega_{0}t+\phi_{0})\right] \nonumber\\
&  +r_{0}\widehat{j}\left[  \sin(\omega_{0}t+\phi_{0})-\frac{eq\sin(\omega
_{0}t+\phi_{0})}{\omega_{0}^{2}mr_{0}x_{q}^{2}}\cos\left(  \omega_{0}%
t+\phi_{0}\right)  \right]  \label{ee7}%
\end{align}
where we have used the approximations for small $\delta,$ $\sin(\theta
+\delta)\approx\sin\theta+\delta\cos\theta$ and $\cos(\theta+\delta
)\approx\cos\theta-\delta\sin\theta.$ \ 

The first thing which we notice is that the interaction of the magnetic moment
(fixed-path model) has led to a nonrelativistic (zero-order in $v/c)$ average
electric dipole moment $\left\langle \overrightarrow{\mathfrak{p}%
}\right\rangle .$\ \ Thus the average electric dipole moment follows from Eq.
(\ref{ee7}) as%
\begin{equation}
\left\langle \overrightarrow{\mathfrak{p}}\right\rangle =\left\langle
e\mathbf{r}(t)\right\rangle =\widehat{i}\frac{e^{2}q}{2m\omega_{0}^{2}%
x_{q}^{2}}=-\frac{e^{2}}{2m\omega_{0}^{2}}\mathbf{E}_{q} \label{ee8}%
\end{equation}
This looks rather like the polarization found for a charged harmonic
oscillator in the electrostatic field $\mathbf{E}_{q}$ of a charge $q$ except
that the polarization is in the \textit{opposite} direction from the
polarizing electric field. \ Rather than the usual attraction between a charge
and a polarizable material, we find here a repulsion.

There is no average nonrelativistic linear momentum in the circular orbit of
the charge $e$ since $\left\langle \mathbf{p}_{mech\text{ nonrel}%
}\right\rangle =m\left\langle \mathbf{v}\right\rangle =m\left\langle
d\mathbf{r}/dt\right\rangle =0$ using $d\mathbf{r}/dt$ from Eq. (\ref{ee7}).
\ However, there is a net linear momentum at the relativistic order
$v^{2}/c^{2}.$ \ The relativistic expression for mechanical linear momentum of
a particle is $\mathbf{p}_{mech}=m\gamma\mathbf{v}=m\gamma c^{2}%
\mathbf{v}/c^{2}$ where we recognize $m\gamma c^{2}$ as the mechanical energy
(rest energy plus kinetic energy) of the particle. \ The energy conservation
law which goes along with the fixed-path perturbation analysis gives energy
balance for mechanical plus electrostatic potential energy as
\begin{equation}
m\gamma_{0}c^{2}+\frac{eqr_{0}\cos\phi_{0}}{x_{q}^{2}}=m\gamma c^{2}%
+\frac{eqr_{0}\cos\phi}{x_{q}^{2}} \label{ee9}%
\end{equation}
Thus we have%
\begin{equation}
\mathbf{p}_{mech}=m\gamma\mathbf{v}=\frac{m\gamma c^{2}\mathbf{v}}{c^{2}%
}=\left(  m\gamma_{0}c^{2}+\frac{eqr_{0}\cos\phi_{0}}{x_{q}^{2}}-\frac
{eqr_{0}\cos\phi}{x_{q}^{2}}\right)  \frac{1}{c^{2}}\frac{d\mathbf{r}}{dt}
\label{ee10}%
\end{equation}
Then time-averaging using $d\mathbf{r}/dt$ from Eq. (\ref{ee7}) (or indeed
from the unperturbed $\mathbf{v(}t)$ of Eq. (\ref{ee2})), we find%
\begin{equation}
\left\langle \mathbf{p}_{mech}\right\rangle =-\widehat{j}\frac{eq\omega
_{0}r_{0}^{2}}{2x_{q}^{2}c^{2}} \label{ee11}%
\end{equation}
This relativistic mechanical linear momentum is exactly equal in magnitude and
opposite in direction from the electromagnetic linear momentum found in Eq.
(\ref{pp2}). \ Thus it is claimed that the mechanical hidden momentum balances
the electromagnetic field momentum giving a self-consistent, self-contained
system of zero linear momentum with no mystery regarding conservation of
linear momentum. \ Of course, this description of the charge-magnetic-moment
interaction says nothing about the source of the crucial nonrelativistic
(zero-order in $v/c$) forces which constrain the motion of the
current-carrying charge $e$ so that it moves only in a circular path. \ Also,
there is no mention of the unusual electrostatic force between the magnetic
moment and the charge $q$ associated with the nonrelativistic electric dipole
moment in Eq. (\ref{ee8}). \ In the opinion of the present writer, this
fixed-path charge-magnetic-moment interaction description has no more validity
than the interaction description of two point charges $e$ and $q$ which are at
rest at a separation $x_{m}$ and are noted with triumph to have zero linear
momentum. The interaction description remains worthlessly incomplete unless
one accounts for the external nonelectromagnetic forces which are required for
equilibrium or else (if there are no non-electromagnetic forces) one notes the
time evolution of the system under the electromagnetic forces.

\section{Nonrelativistic Perturbation Calculation -- Purely Electromagnetic
Model}

Our second model\cite{comparison} for a magnetic moment involves the charge
$e$ attracted to the massive opposite charge $-e$ by Coulomb attraction. \ In
this case, all the forces are electromagnetic, and there are no additional
nonelectromagnetic forces of constraint which keep the charge $e$ in a
circular orbit. \ Indeed, as pointed out by Solem\cite{Solem} in his article,
\textquotedblleft The Strange Polarization of the Classical
Atom,\textquotedblright\ the interaction of the nonrelativistic hydrogen atom
with an external electric field $\mathbf{E}_{q}$ produces an electric dipole
moment for the magnetic moment which is \textit{perpendicular} to the electric
field $\mathbf{E}_{q}$. \ Qualitatively, the situation is easy to understand.
\ A charge $e$ in a circular Coulomb orbit will be slowed down when moving
toward the external charge $q,$ and therefore the charge $e$ will tend to fall
in closer to the central charge $-e;$ on the other hand, the charge $e$ will
be speeded up when moving away from the external charge $q,$ and therefore the
charge $e$ will tend to move further away from the central charge $-e.$
\ Since the external charge $q$ produces only a small orbital perturbation of
the charge $e$, we expect that the orbit of the charge $e$ will be an
elliptical Coulomb orbit. \ Furthermore, the semi-major axis of the elliptical
orbit will remain essentially unchanged since the charge $e$ is in
approximately periodic motion, moving repeatedly towards and away from the
external charge $q,$ successively losing and gaining kinetic energy from the
field $\mathbf{E}_{q}.$ \ The behavior involving a changing elliptical Coulomb
orbit of fixed semi-major axis is totally different from the
tangential-velocity-changing behavior portrayed in the fixed-path analysis
discussed above. \ The changing ellipticity of the Coulomb orbit leads to a
changing average electric dipole field back at the external charge $q$ which
produces forces on $q$ which are completely different from those predicted by
the fixed-path analysis.

The perturbation analysis of the Coulomb orbit by an external electric field
has been given in other publications.\cite{hydro} \ Here we will sketch the
analysis. \ The displacement $\mathbf{r}$ of the charge $e$ with mass $m$ in a
Coulomb orbit with a charge $-e$ at the origin is given by\cite{check} \
\begin{equation}
\mathbf{r}=\frac{3}{2}\frac{\mathbf{K}}{(-2mH_{0})^{1/2}}+\frac{1}{4H_{0}%
}\frac{d}{dt}[m(\mathbf{r}\times\mathbf{v}\times\mathbf{r}+m\mathbf{v}r^{2}]
\label{kk1}%
\end{equation}
where $\mathbf{K}$ is the Laplace-Runge-Lenz vector\cite{Goldstein}%
\begin{equation}
\mathbf{K}=\frac{1}{(-2mH_{0})^{1/2}}\left(  [\mathbf{r}\times(m\mathbf{v}%
)]\times(m\mathbf{v})+me^{2}\frac{\mathbf{r}}{r}\right)  \label{kk2}%
\end{equation}
and $H_{0}$ is the energy of the particle%
\begin{equation}
H_{0}=\frac{1}{2}mv^{2}-\frac{e^{2}}{r} \label{kk3}%
\end{equation}
The Laplace-Runge-Lenz vector $\mathbf{K}$ is constant in time for a Coulomb
orbit. \ Thus the time-average value of $\mathbf{r}$ averaged over the orbit
follows from Eq. (\ref{kk1}) as
\begin{equation}
\left\langle \mathbf{r}\right\rangle =\frac{3}{2}\frac{\mathbf{K}}%
{(-2mH_{0})^{1/2}} \label{kk4r}%
\end{equation}
The nonrelativistic equation of motion for the charge $e$ in the presence of
the perturbing electric field $\mathbf{E}_{q}$ is
\begin{equation}
m\frac{d^{2}\mathbf{r}}{dt^{2}}=-\frac{e^{2}\mathbf{r}}{r^{3}}+e\mathbf{E}_{q}
\label{kk5}%
\end{equation}
Then the time-derivative of $\mathbf{K}$ follows from Eqs. (\ref{kk2}) and
(\ref{kk5}) as
\begin{equation}
\frac{d\mathbf{K}}{dt}=me[-2\mathbf{r}(\mathbf{v}\cdot\mathbf{E}%
_{q})+\mathbf{E}_{q}(\mathbf{r}\cdot\mathbf{v})+\mathbf{v}(\mathbf{r}%
\cdot\mathbf{E}_{q})] \label{kk6}%
\end{equation}
Since the right-hand side of Eq. (\ref{kk6}) is already first order in the
perturbing field $\mathbf{E}_{q}$, we may evaluate the time derivative through
first order by averaging the right-hand side over an unperturbed Coulomb orbit
to obtain%
\begin{equation}
\frac{d\mathbf{K}}{dt}=\frac{3}{2}me[(\mathbf{r}\times\mathbf{v}%
)\times\mathbf{E}_{q}=\frac{3}{2}e\mathbf{L}\times\mathbf{E}_{q}%
=3cm\overrightarrow{\mu}\times\mathbf{E}_{q} \label{kk7}%
\end{equation}
where $\mathbf{L}=m\mathbf{r}\times\mathbf{v}$ is the orbital angular momentum
of the charge $e.~\ \ $Thus in the Coulomb-orbit model for a magnetic moment,
the current-carrying charge $e$ may start out in a circular orbit where the
Laplace-Runge-Lenz vector $\mathbf{K}$ is zero, but the orbit changes in time
becoming increasingly elliptical. \ The magnetic moment $\overrightarrow{\mu}$
changes in time along with the orbital angular momentum $\mathbf{L}$%
\begin{align}
\left\langle \frac{d\overrightarrow{\mu}}{dt}\right\rangle  &  =\frac{e}%
{2mc}\left\langle \frac{d\mathbf{L}}{dt}\right\rangle =\frac{e}{2mc}%
\left\langle \overrightarrow{\mathbf{\Gamma}}\right\rangle =\frac{e}%
{2mc}\left\langle e\mathbf{r}\right\rangle \times\mathbf{E}_{q}\nonumber\\
&  =\frac{e^{2}}{2mc}\frac{3}{2}\frac{\mathbf{K\times E}_{q}}{(-2mH_{0}%
)^{1/2}} \label{kk8}%
\end{align}
The changing orbit of the charge $e$ in the magnetic moment will lead to
electrical forces back on the external charge $q,$ both nonrelativistic
electrostatic forces associated with the presence of the electric dipole
moment and also relativistic forces associate with the electric field induced
by the changing magnetic moment. \ It has been pointed out that the forces
associated with this changing magnetic moment are qualitatively appropriate to
account for the Aharonov-Bohm phase shift as a lag effect associated with
classical electromagnetic forces.\cite{b2006}

\subsection{Closing Summary}

The problem of the interaction of a point charge and a magnetic moment is an
old problem surrounded by controversy. \ In recent years, one version
(involving hidden momentum) of this controversy has been introduced into the
textbook literature of electromagnetism. \ In this article, we present two
models for magnetic moments and note their contrasting behaviors in the
presence of an external charged particle. \ On the one hand, the fixed-path
model for the magnetic moment indeed exhibits hidden momentum, but it involves
unmentioned nonelectromagnetic forces which are of nonrelativistic order and
are vastly greater than the relativistic mechanical effects which are touted
in the textbook literature. \ The fixed-path model also involves an unusual
nonrelativistic electric dipole moment. \ On the other hand, the Coulomb-orbit
model involves only electromagnetic interactions. \ This electromagnetic model
shows a changing magnetic moment which introduces both electrostatic fields
and induced electric fields. \ We believe that the interaction of a charged
particle and a magnetic moment (appropriate for describing Nature) remains a
poorly-understood aspect of electromagnetic theory and that it is premature to
accept the hidden-momentum description of the interaction.


\bigskip

\bigskip

\begin{thebibliography}{99}                                                                                               %


\bibitem {Griffiths}See, for example, D. J. Griffiths, \textit{Introduction to
Electrodynamics} 3rd edn (Prentice-Hall, Upper Saddle River,NJ 1999).

\bibitem {Jackson}See, for example, J. D. Jackson, \textit{Classical
Electrodynamics} 3rd edn (Wiley, New York 1999).

\bibitem {Shockley}W. Shockley and R. P. James, \textquotedblleft`Try simplest
cases' discovery of `hidden momentum forces' on `magnetic
currents',\textquotedblright\ Phys. Rev. Lett. \textbf{18}, 876-879 (1967).

\bibitem {Vaidman}L. Vaidman, \textquotedblleft Torque and force on a magnetic
dipole,\textquotedblright\ Am. J. Phys. \textbf{58}, 978-983 (1990).

\bibitem {RLetter}D. J. Griffiths, \textquotedblleft Resource Letter EM-1:
Electromagnetic Momentum,\textquotedblright\ Am. J. Phys. \textbf{80}, 7-18
(2012), \textquotedblleft Section C. Hidden Momentum.\textquotedblright\ 

\bibitem {AB}Y. Aharonov and D. Bohm, \textquotedblleft Significance of
electromagnetic potentials in quantum theory,\textquotedblright\ Phys. Rev.
\textbf{115}, 485-491 (1959).

\bibitem {AC}Y. Aharonov and A. Casher, \textquotedblleft Topological quantum
effects for neutral particles,\textquotedblright\ Phys. Rev. Lett.
\textbf{53}, 319-321 (1984).

\bibitem {Griffiths2}See, for example, Ref. 1, pp. 357, 361, 520-521.

\bibitem {Jackson2}See, for example, Ref. 2, pp. 189, 618.

\bibitem {objection}See, for example, T. H. Boyer, \textquotedblleft
Relativity, energy flow, and hidden momentum,\textquotedblright\ Am. J. Phys.
\textbf{73}, 1184-1185 (2005); \textquotedblleft Connecting linear momentum
and energy for electromagnetic systems,\textquotedblright\ Am. J. Phys.
\textbf{74}, 742-743 (2006); \textquotedblleft Interaction of a Point Charge
and a Magnet: Comments on `Hidden Mechanical Momentum Due to Hidden
Nonelectromagnetic Forces',\textquotedblright\ arXiv:0708.3367v1 (2007);
\textquotedblleft Concerning `hidden momentum',\textquotedblright\ Am. J.
Phys. \textbf{76}, 190-191 (2008).

\bibitem {refuse}The editors of the Physical Review A refused to send out for
review manuscripts such as T. H. Boyer, \textquotedblleft Classical
Electromagnetic Interaction of a Point Charge and a Magnetic Moment:
Considerations Related to the Aharonov-Bohm Phase Shift,\textquotedblright%
\ Found. Phys. \textbf{32}, 1-39 (2002).

\bibitem {Mansuripur}M. Mansuripur, \textquotedblleft Trouble with the Lorentz
law of force: Incompatibility with special relativity and momentum
conservation,\textquotedblright\ Phys. Rev. Lett. \textbf{108}, 193901 (2012).

\bibitem {Jackson3}See, for example, Ref.2, p. 186.

\bibitem {comparison}The basic \textquotedblleft fixed-path\textquotedblright%
\ model appears as an example, in Ref. 1, p. 520-521. \ The \textquotedblleft
fixed-path\textquotedblright\ circular orbit is discussed in the arXiv article
in Ref. 10. \ The Coulomb-force model appears in Ref. 11.

\bibitem {Darwin}See ref. 2, pp. 596-598.

\bibitem {PageandAdams}See, for example, L. Page and N. I. Adams,
\textquotedblleft Action and reaction between moving
charges,\textquotedblright\ Am. J. Phys. \textbf{13}, 141--147 (1945).

\bibitem {Solem}J. C. Solem, \textquotedblleft The strange polarization of the
classical atom,\textquotedblright\ Am. J. Phys. \textbf{55}, 906-909 (1987).

\bibitem {hydro}See refs. 11 and 17. \ See also, L. C. Biedenharn, L. S.
Brown, and J. C. Solem, \textquotedblleft Comment on \ `The strange
polarizationof the classical atom',\textquotedblright\ Am. J. Phys.
\textbf{56}, 661-663 (1988),\ and J. C. Solem, \textquotedblleft Variations on
the Kepler problem,\textquotedblright\ Found Phys. \textbf{27}, 1291-1306 (1997).

\bibitem {check}Equation (\ref{kk1}) can be check by carrying out the time
derivative and inserting the equation of motion $\mathbf{a}=-e^{2}%
\mathbf{r}/(mr^{3})$.

\bibitem {Goldstein}H. Goldstein, C. Poole, J. Safko, \textit{Classical
Mechanics} 3rd edn (AddisonWesley, New York 2002), pp. 103-106.

\bibitem {b2006}T. H. Boyer, \textquotedblleft Darwin-Lagrangian analysis for
the interaction of a point charge and a magnet: Considerations related to the
controversy regarding the Aharonov-Bohm and Aharonov-Casher phase
shifts,\textquotedblright\ J. Phys. A: Math. Gen. \textbf{39}, 3455-3477 (2006).
\end{thebibliography}
\end{document}